\documentclass[10pt,aps, pra, reprint, footinbib]{revtex4-1}

\usepackage[T1]{fontenc}
\usepackage[english]{babel}
\usepackage{graphicx}
\usepackage{amsmath}
\usepackage{amssymb}

\begin{document}

\title{Light-intensity ssusceptibility and ``active'' noise spectroscopy}
\author{G.G.Kozlov}
\email{gkozlov@photonics.phys.spbu.ru}
\author{V.S.Zapasskii}
\email{zap@vz4943.spb.edu} \affiliation{Spin Optics Laboratory and
Physics Department, Saint-Petersburg State University, Saint
Petersburg, 198504 Russia}
\date{\today}

\begin{abstract}
In this paper, we consider informative potentialities of the
``active'' optical noise spectroscopy, under which we understand,
generally, spectroscopy of response of a multilevel quantum system
to the resonant optical field with its intensity modulated by
``white'' noise. We show that calculations of such a response can
be most conveniently performed, in the linear approximation, by
introducing the notion of light-intensity susceptibility (LIS)
whose spectrum is determined by Laplace transform of the response
to a small step-wise change of the optical field intensity. The
results of calculations for a simple four-level quantum system
show that its LIS spectrum may provide information not only about
the ground-state structure (like conventional
Faraday-rotation-based spin noise spectroscopy), but also about
properties of the optical transitions (including nutation
frequencies in the applied optical field). From the experimental
point of view, such a noise spectroscopy of the intensity-related
susceptibility  can be especially efficient in combination with
the up-to-date spectrum analyzers providing extremely fast data
processing.
\end{abstract}

\maketitle

\section{Introduction}

Optical spectroscopy, with its frequencies lying in the range of $10^{15}$ Hz, is known to be capable
of studying spectral features of the system at frequencies many orders of magnitude lower (like, e.g.,
Zeeman or hyperfine splittings). Such a high-resolution spectroscopy often employs, for detection of
these features, the intensity correlation characteristics of the light coupled to the system. Ideology of
the light intensity noise spectroscopy has been primarily realized by Forrester \cite{Forrester} and
Hunbury-Brown and Twiss \cite{HBT} and then has been further developed for studying dynamics of motion
of different atomic and molecular systems including dynamics of nonstationary quantum states (see,
e.g., \cite{Aleksandrov}). With the advent of lasers, the light intensity noise spectroscopy has crucially
increased its sensitivity and gained a deeper practical sense.

The methods of noise spectroscopy intended for studying spectra of eigen-frequencies of quantum systems
can be divided into two main classes. In  the measurements of the first class, optical probing of
the sample is supposed to be nonperturbative: the light beam passing through the sample in the region
of its transparency acquires additional intensity or polarization fluctuations that carry the sought
information. Ideally, intrinsic noise of the probe beam, in this approach, is negligibly small. A
typical example of the experimental technique of this class is the Faraday-rotation-based spin noise
spectroscopy (SNS), which provides a unique opportunity of studying magnetic resonance and spin dynamics
of atomic and solid-state systems practically in a perturbation-free way
\cite{Zap,Crooker3,Oestreich,Zap1}. The
recent enormous growth of interest to this experimental technique, primarily demonstrated more than 30
years ago \cite{Zap}, is associated, to a considerable extent, with great progress in the present-day
electronics, which has made it possible to digitize arrays of data with the sampling rates up to
several GHz and to perform fast Fourier transform (FFT) in real time. As a result, the new class of
spectrum analyzers with FFT has allowed to increase sensitivity of the measurements and,
correspondingly, to shorten the acquizition time by several orders of magnitude (as compared to the
standard sweeping spectral analyzers). Due to this technical refinement, SNS has been successfully
applied to semiconductor systems, including microsamples and nanostructures \cite{Oestreich} and is now
about to turn into a routine tool of magnetic spectroscopy.

The other experimental approach to the noise spectroscopy that has been developed in 90s and applied to
atomic systems, is based on the use of a randomly modulated light beam acting on the system in the
region of its absorption \cite{Yab,McIntyre,Walser,Mitsui,Martinelly,Valente}. This technique is
deliberately perturbative: the system excited by the pump light modulated in a wide frequency range
responds selectively to certain resonant frequencies, and, as a result, the noise spectrum of the
transmitted light appears to be modified. In the above papers, as a source of light was used a diode
laser with specific correlation properties of its output emission: intensity fluctuations were
extremely low (they were claimed to lie below the shot-noise level \cite{Machida}), whereas its
frequency noise was well pronounced. The frequency modulated light, due to interaction with a narrow
spectral line of the atomic medium, produced the required noisy excitation of the system. It was shown,
in those experiments that the atomic system, under certain conditions, may exhibit resonant response
not only at the frequency of the ground-state magnetic resonance (Larmor precession), as it was
primarily demonstrated by Bell and Blum\cite{B&B}, but also at Rabi frequency of the resonant
transition\cite{Walser}, which may substantially widen informative potential of the noise spectroscopy.

It is noteworthy that this kind of noise spectroscopy was never
considered as a possible experimental tool for studying a wider
circle of materials, including semiconductor systems and
nanostructures, highly important for contemporary physics and
applications. At the same time, from the methodological viewpoint,
experimental measurements of the intensity-related susceptibility
using fluctuating optical fields have much in common with the
technique of conventional ``passive'' noise spectroscopy and
thereby have all chances to take an advantage of the new
instrumental opportunities provided by the up-to-date electronics.

In this paper, we consider a general theoretical model that describes the effect of a four-level system
on the intensity noise spectrum of the transmitted resonant light. The light intensity noise spectrum
is assumed to be "white" and essentially exceeding in magnitude the shot-noise level. We analyze
interaction of the light beam with the medium by introducing the notion of the {\it light-intensity
susceptibility} (LIS), which in our opinion, may be useful for solving certain problems of nonlinear optics.
In particular, this notion was, in fact, used in the analysis of dynamics of a saturable absorber in
modulated optical fields \cite{Slowlight}.

In what follows we study spectral behavior of the intensity-related susceptibility in the range of
relatively low frequencies (e.g., of the ESR range) . We will show that frequency dependence of the
light-intensity susceptibility  contains information not only about energy and relaxation
characteristics of the system accessible for linear spectroscopy, (including conventional "passive"
spin noise spectroscopy), but also about specific dynamics of optical transitions (nutation
frequencies) in the field of the noisy resonant pumping. We will see that, in contrast to the
conventional linear optical spectroscopy, which allows one to measure frequencies of transitions
between the levels of unperturbed states of the system, the LIS spectroscopy
  makes it possible (for sufficiently long relaxation times) to measure frequencies of
transitions between the states of some effective Hamiltonian dependent on the optical field intensity
 $I_0$.\cite{footnote_1}

\section{Starting Points}

We define the {\it light-intensity susceptibility} (LIS)
  in the following way. Assume that the system under study
is subjected to action of a quasi-monochromatic optical field of the frequency $\omega$, with its
intensity $I$ having a dc ($I_0$) and weak ac ($\delta I(t)$) components $I=I_0+\delta I(t)$.
 Correspondingly, the power $P$ absorbed by the
system will also have the dc and ac components: $P=P_0+\delta P(t)$.
 We define the light-intensity susceptibility  as a
linear susceptibility $ \chi(\imath\xi)$
 connecting the ac components of the intensity and absorption in the case when
$\delta I(t)\rightarrow \delta I e^{-\imath\xi t}, \delta P(t)\rightarrow \delta P e^{-\imath\xi t}$
 and $\delta P=\chi(\imath\xi)\delta I$.

\begin{figure}
\includegraphics[width=\columnwidth,clip]{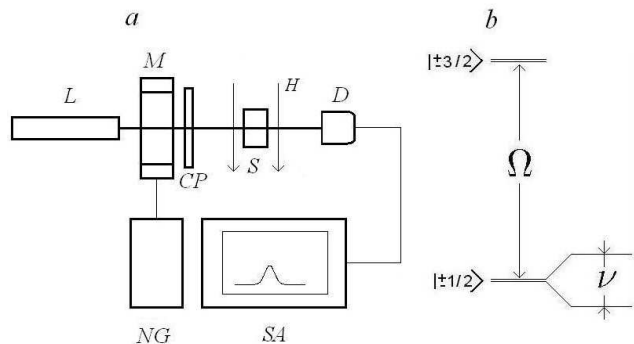}
\caption{(a)~schematic of experimental setup for observation of
light intensity susceptibility. L -- laser, M -- optical
modulator, CP -- circular polarizer, S -- sample, D --
photodetector, SA -- spectrum analyzer, NG -- noise generator.
(b)~energy structure of the model system.}\label{fig_One}
\end{figure}

A possible experimental observation of frequency dependence
(spectrum) of the light-intensity susceptibility with the use of
the spectrum analyzer looks as follows [Fig.~\ref{fig_One}(a)].
The laser beam subjected to a broadband noise intensity modulation
\cite{footnote_2}
passes through the sample under study and hits the
photodetector, whose output signal is fed to the input of the
spectrum analyzer. If the intensity modulation spectrum of the
laser beam is "white" , then we will observe, at the output of the
spectrum analyzer, the LIS spectrum
 of the studied sample. The experiments of this kind have been
described in \cite{Yab,Mitsui} where the resonant paramagnetic susceptibility of atomic gases has been detected.
It should be emphasized that, though we consider a linear intensity susceptibility, it can be
revealed only by optically nonlinear media. In the next section, we will calculate the LIS of the
simplest model system and will study its spectrum.

 \section{Model Calculations }

The sample under study is assumed to be an ensemble of particles,
and the total power absorbed by the sample irradiated by the
electromagnetic field to be equal to the sum of contributions of
individual particles. The energy spectrum of individual particle
is assumed to consist of two doublets [Fig.~\ref{fig_One}(b)]: the
ground $|\pm 1/2\rangle$ and the excited $|\pm 3/2\rangle$ one
 (here we indicate projection of the angular moment onto the light propagation
direction, which is assumed to coincide with the z-axis of the our
coordinate system). This energy structure arises, in particular,
in the atomic multiplet perturbed by an axially symmetric field.
\cite{footnote_3}
In this case, magnetic splitting of the
ground doublet is characterized by isotropic $g$-factor, whereas
the doublet
 $|\pm 3/2\rangle$ is split only by
the field parallel to  the $z$-axis. Spectral distance between the doublets (denoted by $\Omega$) is
supposed to be close to frequency $\omega$
 of the optical  field $\omega\approx\Omega$ and essentially higher than other frequency
parameters of the problem.

We will calculate the LIS spectrum we are interesting in as follows.  Let us assume that the
system under study was subjected, during sufficiently long
time, to a monochromatic optical field with fixed intensity, so that the
power absorbed by the system has already reached its steady-state value.
Let the amplitude of this field now experience a jump that induces a transient
process corresponding to transition to a new stationary state. Below, we will find
the transient characteristics (i.e., temporal behavior of the absorbed power during this process),
which is connected with the sought susceptibility by the Laplace transform.

Let us pass to implementation of the above program for the
conditions of the experiment \cite{B&B}. The optical field acting
upon the system is assumed to be circularly polarized and the
whole system to be placed in a transverse magnetic field $\cal H$
directed along the $x$-axis [Fig.~\ref{fig_One}(a)].

Since the magnetic splitting of the ground doublet is supposed  to be small compared with the optical
frequency $\omega$, among the matrix elements of the operator of interaction of the system with
optical field (denote it $H_1$), we have to take into account only those corresponding to quasi-resonant
transitions:

$$
\begin{cases}
\langle -1/2 |H_1|+1/2\rangle=\hbox{ non resonant}\cr \langle +1/2
|H_1|-1/2\rangle=\hbox{ non resonant}\cr \langle +3/2
|H_1|+1/2\rangle= A e^{-\imath\omega t}\cr \langle -3/2
|H_1|-1/2\rangle= A e^{\imath\omega t}\cr \langle +1/2
|H_1|+3/2\rangle= A e^{\imath\omega t}\cr \langle -1/2
|H_1|-3/2\rangle= A e^{-\imath\omega t}
\end{cases}
$$
 Here, the constant $A$ describes intensity of the transitions under circularly polarized excitation. If
we arrange possible states of the system in the order
 $|+1/2\rangle, |-1/2\rangle, |-3/2\rangle, |+3/2\rangle$, then the matrix $H_1$ will have the form
 \begin{equation}
 H_1=A\begin{pmatrix}
 0 & 0 & 0 &\mu\cr 0 & 0 & \bar\mu & 0\cr
 0 & \mu & 0 & 0\cr \bar\mu & 0 &0 &0\end{pmatrix}
 \hskip10mm \mu\equiv e^{\imath\omega t}
\label{2}
 \end{equation}
At $A=0$, the Hamiltonian matrix of the system, in the basis introduced above, will have the form
\begin{equation}
H_0=\begin{pmatrix}0 & \nu & 0 & 0\cr \nu & 0 & 0 & 0\cr
 0 & 0 & \Omega & 0\cr 0 & 0 &0 &\Omega\end{pmatrix}\hskip10mm \nu\equiv {g\beta {\cal H}\over 2
 \hbar}
\label{3}
\end{equation}
At $A \ne 0$, the Hamiltonian of the system depends on time and equals $H=H_0+H_1$. Neglecting the relaxation
processes (they will be taken into account later), dynamics of the density matrix of the system $\rho$ can
be described by the Liouville equation  $\imath\dot\rho=[H,\rho]$.
 To treat this equation, it is convenient to pass to the
basis,  in which $H_0$ is diagonal, and then to the "rotating coordinate frame", where the total
Hamiltonian is time-independent. The first step is made using a unitary transformation with the matrix
 \begin{equation}
  S\equiv{1\over \sqrt 2}\begin{pmatrix}1&1&0&0\cr 1&-1&0&0\cr0&0&\sqrt 2&0\cr 0&0&0&\sqrt
  2\end{pmatrix},\hskip2mm\hbox{ with } S^{-1}=S
\label{4}
  \end{equation}

We will mark with tilde
 all the operators having the form  $\tilde H\equiv SHS$. Then, direct calculations show that
\begin{equation}
\tilde H_0=\begin{pmatrix}\nu&0&0&0\cr 0&-\nu&0&0\cr 0&0&\Omega
&0\cr 0&0&0&\Omega\end{pmatrix} \label{5}
\end{equation}
\begin{equation}
\tilde H_1={A\over \sqrt 2}
\begin{pmatrix}0&0&\bar\mu&\mu\cr 0&0&-\bar\mu &\mu\cr \mu&-\mu&0 &0\cr \bar\mu&\bar\mu&0&0\end{pmatrix}
\label{6}
\end{equation}
And the density matrix $\tilde \rho=S\rho S$ meets the equation  $\imath\partial\tilde\rho/\partial t
=[\tilde H,\tilde\rho]$.
 Transition to the "rotating coordinate frame" is performed using the following time-dependent
transformation of the operators. Let us pass to the hatted operators and new density matrix using the
rule
 \begin{equation}
\widehat H=e^{\imath Mt} \tilde H e^{-\imath Mt}\hskip5mm \sigma\equiv e^{\imath Mt}\tilde \rho e^{-\imath Mt}
\label{7}
\end{equation}
Where the matrix $M$ has the following elements
\begin{equation}
M\equiv\begin{pmatrix}0&0&0&0\cr 0&0&0&0\cr 0&0&\omega &0\cr
0&0&0&\omega\end{pmatrix} \label{8}
\end{equation}
Then, one can easily see that $\sigma$ satisfies the equation
 $\imath\dot\sigma=[\widehat H-M,\sigma]$. Calculations of the matrix $\widehat H$ show
that it contains time-independent elements and
  elements proportional to  $e^{\pm 2\imath\omega t}$.
 The matrix elements
oscillating at double frequency are ignored as essentially nonresonant and not affecting dynamics of
the density matrix $\sigma$. Thus, the matrix $\widehat H$,
 with allowance for the aforesaid, can be written in the
form:
 \begin{equation}
 \widehat H=\begin{pmatrix}\nu &0&0& d\cr 0&-\nu &0 &d\cr 0&0&\Omega&0\cr d &d & 0&\Omega\end{pmatrix}
\hskip10mm d\equiv {A\over \sqrt 2}
 \label{9}
 \end{equation}
It follows from this expression that the third column and third row of the matrix $H$ are mutually
diagonal and can be omitted. Now, all the operators can be presented by 3$\times$3 matrices, and the equation
of motion for the matrix $\sigma$ acquires the form
\begin{equation}
\imath\dot\sigma=[W,\sigma],\hskip 10mm W\equiv\begin{pmatrix}\nu
& 0& d \cr 0 & -\nu & d \cr d & d & \Delta\end{pmatrix},
\hskip10mm \Delta\equiv \Omega-\omega \label{10}
\end{equation}
with the time-independent matrix $W$.

As mentioned above, our task is to calculate transient dynamics of the power $P$ absorbed by the system
after a jump of the optical field amplitude. This power can be expressed in the form
\begin{equation}
P={d\over d t}\hbox{ Sp } \rho H={d\over d t}\hbox{ Sp } \sigma \widehat H=
\hbox{ Sp } \dot\sigma \widehat H +\hbox{ Sp } \sigma {d \widehat H\over d t}
\label{11}
\end{equation}
The last term in (\ref{11}) can be neglected because the operator  $\widehat H$
 oscillates in time at double optical
frequency, and the mean value of the corresponding contribution vanishes. Thus,
\begin{equation}
P= \hbox{ Sp } \dot\sigma \widehat H =-\imath\hbox{ Sp }[\widehat H-M,\sigma]\widehat H=
\imath\hbox{ Sp }[M,\sigma]\widehat H
\label{12}
\end{equation}

Using explicit formulas (8) and (9) for the matrices $M$ and  $\widehat H$, we finally have
\begin{equation}
P=2\omega d\bigg[\hbox{ Im }\sigma_{13}+\hbox{ Im }\sigma_{23}\bigg]
\label{13}
\end{equation}

Note that account of relaxation terms in the equation of motion for the density matrix leads to
appearance of terms proportional to the relaxation rates in the expression for the absorbed power $P$.
We will assume that these rates are much lower than the optical frequency $\omega$ entering Eq (\ref{13})
and will use Eq. (\ref{13}) in the presence of relaxation processes also.

Now, we will write down Eq. (\ref{10})
 in components of the density matrix and introduce relaxation needed to
obtain the steady-state regime of absorption:
\begin{equation}
\begin{cases}
\imath\dot\sigma_{12}=\sigma_{12}[2\nu-\imath
T_{12}^{-1}]+d[\sigma_{32}-\sigma_{13}] \cr
\imath\dot\sigma_{21}=-\sigma_{21}[2\nu+\imath
T_{21}^{-1}]+d[\sigma_{31}-\sigma_{23}]\cr
\imath\dot\sigma_{13}=\sigma_{13}[\nu- \Delta-\imath
T_{13}^{-1}]+d[\sigma_{33}-\sigma_{11}-\sigma_{12}]\cr
\imath\dot\sigma_{31}=\sigma_{31}[\Delta-\nu+\imath
T_{31}^{-1}]+d[\sigma_{11}+\sigma_{21}-\sigma_{33}]\cr
\imath\dot\sigma_{23}=\sigma_{23}[-\Delta-\nu-\imath
T_{23}^{-1}]+d[\sigma_{33}-\sigma_{21}-\sigma_{22}]\cr
\imath\dot\sigma_{32}=\sigma_{32}[\Delta+\nu+\imath
T_{32}^{-1}]+d[\sigma_{22}-\sigma_{33}+\sigma_{12}]\cr \imath
\dot\sigma_{11}=d[\sigma_{31}-\sigma_{13}]+\sigma_{33}\tau_1^{-1}+(\sigma_{22}-\sigma_{11})T_1^{-1}\cr
\imath
\dot\sigma_{22}=d[\sigma_{32}-\sigma_{23}]+\sigma_{33}\tau_1^{-1}-(\sigma_{22}-\sigma_{11})T_1^{-1}\cr
\imath\dot\sigma_{33}=d[\sigma_{13}-\sigma_{31}+\sigma_{23}-\sigma_{32}]-2\sigma_{33}\tau_1^{-1}\hskip40mm
\end{cases}
\label{14}
\end{equation}
The form of the relaxation terms for diagonal elements of the
density matrix implies that states 1 and 2 are the lowest, and the
temperature is so high that they are equally populated. State 3,
on the contrary, is assumed to lie so high that its equilibrium
population is zero. These assumptions correspond to the conditions
$g\beta {\cal H}<<kT$ and $\hbar\Omega>>kT$.
 Let us introduce the following notations for the real and imaginary
parts of the matrix elements $\sigma$ and relaxation times
\begin{subequations}
\begin{align}
&\begin{cases} \sigma_{12}+\sigma_{21}=x_1\cr
\sigma_{13}+\sigma_{31}=x_2\cr
 \sigma_{23}+\sigma_{32}=x_3
\end{cases}
\\
&\begin{cases} \sigma_{12}-\sigma_{21}=\imath y_1\cr
\sigma_{13}-\sigma_{31}=\imath y_2\cr
\sigma_{23}-\sigma_{32}=\imath y_3
\end{cases}
\\
&\begin{cases} \sigma_{11}=s_1\cr \sigma_{22}=s_2\cr
\sigma_{33}=s_3
\end{cases}
\\
&\begin{cases} T_{12}=T_{21}=T_2\cr T_{13}=T_{31}=\tau_2\cr
T_{23}=T_{32}=\tau_2
\end{cases}
\label{15}
\end{align}
\end{subequations}
Here, we denoted the dephasing time in the magnetic doublet as
$T_2$ and the dephasing time in the transitions 2-3 and 1-3,
saturated by the optical field, as $\tau_2$. By defining the
vector-column $v$ as
\begin{equation}
v=\begin{pmatrix}x_1\cr x_2\cr x_3\cr y_1\cr y_2\cr y_3\cr s_1\cr
s_2\cr s_3\end{pmatrix} \label{17}
\end{equation}
 and the matrix $G(d)$ as
\begin{widetext}
\begin{equation}
G(d)=\begin{pmatrix}-T_2^{-1} & 0 & 0 & 2\nu & -d & -d & 0 & 0 &
0\cr 0 & -\tau_2^{-1} & 0 & -d & -\Delta+\nu & 0 & 0 & 0 & 0\cr 0
& 0 & -\tau_2^{-1} & d & 0 & -\Delta-\nu & 0 & 0 & 0\cr -2\nu & d
& -d & -T_2^{-1} & 0 & 0 & 0 & 0 & 0\cr d & \Delta-\nu & 0 & 0 &
-\tau_2^{-1} & 0 & 2d & 0 & -2d\cr d & 0 & \Delta+ \nu &0 & 0 &
-\tau_2^{-1} & 0 & 2d & -2d\cr 0 & 0 & 0 & 0 & -d & 0 & -T_1^{-1}
& T_1^{-1} &\tau_1^{-1}\cr 0 & 0 & 0 & 0 & 0 & -d & T_1^{-1} &
-T_1^{-1} &\tau_1^{-1}\cr 0 & 0 & 0 & 0 & d & d & 0 & 0
&-2\tau_1^{-1} \end{pmatrix}, \label{18}
\end{equation}
\end{widetext}
Using Eq.(\ref{14}), we obtain the following
equation for $v$
\begin{equation}
{dv\over dt}= G(d)v
\label{19}
\end{equation}
Since the last row of matrix $G$ is the sum of two previous ones, this matrix should always have a zero
eigen-number. The corresponding eigen-vector (denote it $e^9$) allows us to obtain the steady-state
regime of the system saturated by the optical field. The vector  $b$,
 corresponding to the steady-state
matrix of the system, evidently represents the vector $e^9$ normalized as $b=q^{-1}e^9$,  where
  $q\equiv e^9_7+e^9_8+e^9_9$. In accordance with the
formulation of the problem, given above, we assume that the density matrix of the system, before the
jump of the optical field amplitude, was determined by the vector $b$. In this case, the power $P_0$
absorbed by the system before the jump can be expressed through the components of this vector using Eq.
(\ref{13})
\begin{equation}
P_0=\omega d [b_5+b_6]
\label{20}
\end{equation}
If the jump of the optical field amplitude  $d\rightarrow d+\delta d$ occurs at $t=0$,
 then, to find dynamics of the system after
the jump, we have to solve the equation  $\dot v=G(d+\delta d)v$
  with the initial condition $v(t=0)=b$. Denote by $p^i, i=1,2,...,9$ the
eigen-vectors of the matrix $G(d+\delta d)$ and decompose the vector $b$ over them:
\begin{equation}
b=\sum_i C_ip^i,
\label{21}
\end{equation}
Vectors $p^i$ and coefficients $C_i$ can be easily found numerically. Now, the vector $v$
  describing dynamics
of the system after the jump can be written in the form
\begin{equation}
v=\sum_i C_i p^i e^{\lambda_i t},
\label{22}
\end{equation}
Where $\lambda_i$ are the eigen-numbers of the matrix $G(d+\delta d)$.
 Dynamics of the absorbed power is given by the equation
\begin{equation}
P(t)=\omega(d+\delta d)\sum_{i=1}^9 C_i e^{\lambda_i t}[p^i_5+p^i_6]
\label{23}
\end{equation}

The quantity $\delta P\equiv P(t)-P_0$ can be considered as a
response of the system to the jump $\delta d$, and, at small $\delta d$, this
response should be linear. The LIS we are interested in, is evidently proportional to the
susceptibility $K(\imath\xi)$ connecting $\delta P$ and $\delta d$.
 Then, if $S(\xi)$ is the Laplace transform of the response to the
jump,
\begin{widetext}
 \begin{equation}
  S(\xi)=\lim_{\alpha\rightarrow +0}\int_0^\infty e^{\imath\xi t}[P(t)-P_0] e^{-\alpha t}dt=
-\omega (d+\delta d)
\sum_{i=1}^9\bigg[{C_i(p_5^i+p_6^i)\over \imath\xi+\lambda_i}\bigg]
+{\omega d (b_5+b_6)\over\imath \xi},
  \label{24}
  \end{equation}
\end{widetext}
then  $ |K(\imath\xi)|^2=|\xi S(\xi)|^2$, and, for the LIS
spectrum, we obtain
 \begin{equation}
  |\chi(\imath\xi)|^2\sim|\xi S(\xi)|^2
  \label{25}
  \end{equation}
Thus, calculations of the LIS spectrum, within the simplest model
described above can be performed using Eqs. (\ref{24}) and
(\ref{25}).

\section{Results}

Let us calculate the LIS spectrum  for essentially different relaxation rates  in the optical transitions.

\begin{figure}
\includegraphics[width=.8\columnwidth,clip]{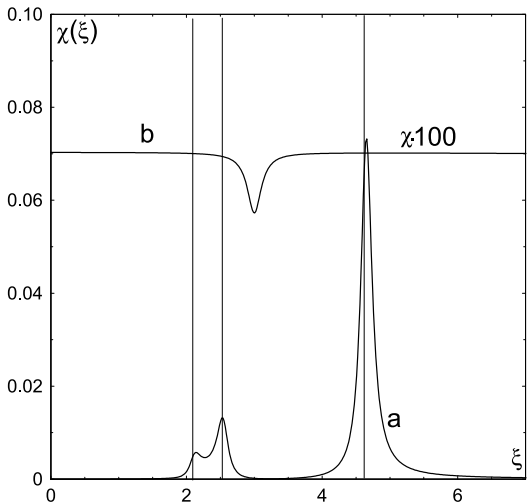}
\caption{(a)~the LIS spectrum for relatively long relaxation times
in the optical transitions
 $\tau_2=10, \tau_1=2\tau_2$,
 (b)~the same for relatively short relaxation times
 $\tau_2=0.01, \tau_1=2\tau_2$. The rest parameters are the same for both figures and are:
$ T_2=10, T_1=2T_2, d=1, \nu=\Delta=1.5$.}\label{fig_Two}
\end{figure}

When the relaxation times entering the equations for elements of the density matrix (\ref{14}) substantially
exceed the period of optical nutation $d$,  the dynamics of the density matrix for times smaller than
the relaxation times is determined by Eq. (\ref{10}). Under these conditions, this dynamics represents
oscillations at frequencies equal to all possible differences between  the eigen-values of the effective
Hamiltonian $W$. The quantities $ W_i, i=1,2,3$, in this case,
can be calculated either numerically or using Cardano's
formula. Thus, the LIS spectrum will reveal peaks at appropriate frequencies, which, in this case,
will be three:  $\xi_1\equiv|W_1-W_2|, \xi_2\equiv|W_1-W_3|,
   \xi_3\equiv|W_2-W_3|$. This is demonstrated in Fig.~\ref{fig_Two}(a), which shows the LIS spectrum obtained using Eqs.
(\ref{24}) and (\ref{25}) for   $ d^{-1}<< T_1, \tau_1, \tau_2$.
 Verical lines indicate frequencies $\xi_i, i=1,2,3$. In the absence of the magnetic field, one
can easily see that
  $ W_{1,2}=[\Delta\pm\sqrt{\Delta^2+4d^2}]/ 2,\hskip10mm W_3=0$.
   At  $d/\Delta<<1$, the LIS spectrum shows peaks at frequencies
   $\xi\approx\Delta$ and $\xi\approx d^2/\Delta$ (the peak is weakly
pronounced), while at  $d/\Delta<<1$, at the frequency of optical nutation $2d$  and at $d$
 (the peak is also weak).

The situation strongly changes when passing to short relaxation
times of the optical transitions $\tau_2,\tau_1<< d^{-1}$. This
case corresponds to the experiment \cite{B&B} when the system
becomes relatively transparent for the circularly polarized light
with its intensity oscillating at the frequency of magnetic
splitting (ESR frequency). In this case, the LIS spectrum shows a
dip at the frequency of magnetic splitting $2\nu$
[Fig.~\ref{fig_Two}(b)]. The condition  $\tau_2,\tau_1<< d^{-1}$
 allows us to neglect the time derivatives
  $\dot\sigma_{13}, \dot\sigma_{31},\dot\sigma_{23},\dot\sigma_{32},\dot\sigma_{33}$ in (\ref{14})
  as compared with the relaxation
terms. This, in turn, makes it possible to express algebraically the matrix elements
   $\sigma_{13}, \sigma_{31}, \sigma_{23}, \sigma_{32}, \sigma_{33}$, pertinent to
optical transitions through the elements
 $\sigma_{12}, \sigma_{21}, \sigma_{11}, \sigma_{22}$,
 related to the magnetically split doublet and, thus, to
obtain the relationships similar to those presented in \cite{B&B}.

\section{Conclusion}

The goal of this paper is to show, using as an example
a four-level system, potential usefulness and
efficiency of the "active" noise spectroscopy that employs
transformation of intensity noise spectrum of the light
passing through the medium as a source of information about
its eigen-frequencies. This type of
the noise spectroscopy has much in common with the
conventional Faraday-rotation-based spin noise
spectroscopy and, under certain conditions,
provides the same information about the system and
also may serve as a method of magnetic spectroscopy.
At the same time, the technique considered here,
though deprived of the advantage
of being nonperturbative, has
a number of specific properties and unquestionable merits.
 In contrast to
the Faraday-rotation-based technique, this method allows
one to get information about optical
transitions and properties of the excited states, rather
than only about the ground-state dynamics and structure.
The signal of the "active" noise spectroscopy, being a
result of induced (rather than spontaneous) response of
the system, can be controlled by the intensity of the
optical field and, evidently, can be easier detected.
This spectroscopy,
unlike the "passive" FR-based, implies pure intensity-related
measurements and, in the proposed version, does not require
any polarimetric sensitivity. This technique employs
fluctuations of the optical field rather than thermodynamic
fluctuations of the ensemble of particles, and, therefore,
the detected signal is not connected explicitely with the
size of the light spot.
It may seem strange, but a serious problem for this kind
of spectroscopy may be the problem of laser source with a
sufficiently high level of intensity noise in a sufficiently
broad frequency range. Perhaps, special efforts will be
needed to create a special laser source for the spectrometer of this kind.
In principle, such a spectroscopy of the light intensity
susceptibility can be implemented with harmonically (rather than randomly)
modulated light. But in this case, to obtain the spectrum,
we will have to scan the modulation frequency, and the
experimental technique will lose its main advantage.

We consider the spectroscopy of the light intensity
susceptibility to be a promising experimental technique
for studying energy structure and dynamics of multilevel
quantum systems under resonant optical excitation.




\begin{references}
\bibitem{footnote_1} For a two-level system, this is the total Hamiltonian of the problem in the
coordinate frame rotating with the optical frequency $\omega$. In
the general case, it can be considered as the Hamiltonian of the
``dressed'' system.
\bibitem{footnote_2} Such a modulation can be realized using a noise generator
and electro-optical modulator, but, however, in some cases,
intrinsic noise of the laser source can be used \cite{Mitsui}
\bibitem{footnote_3} The energy diagram of the system is chosen to be closely
similar to that of a quantum dot in a transverse magnetic field\cite{Shabaev}.
\end{references}
\end{document}